\documentclass[twocolumn,english,showpacs,prl]{revtex4}
\usepackage[T1]{fontenc}
\usepackage[latin9]{inputenc}
\usepackage{color}
\usepackage{amsmath}
\usepackage{amssymb}
\usepackage{graphicx}
\usepackage{esint}

\makeatletter

\providecommand{\tabularnewline}{\\}

\@ifundefined{textcolor}{}
{%
 \definecolor{BLACK}{gray}{0}
 \definecolor{WHITE}{gray}{1}
 \definecolor{RED}{rgb}{1,0,0}
 \definecolor{GREEN}{rgb}{0,1,0}
 \definecolor{BLUE}{rgb}{0,0,1}
 \definecolor{CYAN}{cmyk}{1,0,0,0}
 \definecolor{MAGENTA}{cmyk}{0,1,0,0}
 \definecolor{YELLOW}{cmyk}{0,0,1,0}
}

\usepackage{babel}

\makeatother

\usepackage{babel}
\begin{document}

\title{Hyperuniformity of critical absorbing states}

\author{Daniel Hexner\textsuperscript{1} and Dov Levine\textsuperscript{1,2}}

\affiliation{\textsuperscript{1}Department of Physics, Technion, Haifa 32000,
Israel\linebreak{}
 \textsuperscript{2}Initiative for the Theoretical Sciences - CUNY
Graduate Center \\
365 Fifth Avenue, New York, NY 10016, USA}
\begin{abstract}
The properties of the absorbing states of non-equilibrium models belonging
to the conserved directed percolation universality class are studied.
We find that at the critical point the absorbing states are hyperuniform,
exhibiting anomolously small density fluctuations. The exponent characterizing
the fluctuations is measured numerically, a scaling relation to other
known exponents is suggested, and a new correlation length relating
to this ordering is proposed. These results may have relevance to
photonic band-gap materials.
\end{abstract}


\maketitle
Dynamical systems with absorbing states\cite{Lubek,hinrichsen_non-equilibrium_2000,Non-Equilibrium_Book}
are intrinsically out of equilibrium, as they necessarily violate
detailed balance - by definition, once they arrive at an absorbing
state, they can never get out. There are many such systems, and they have an extensive literature\cite{Non-Equilibrium_Book}.
Recently\cite{pine_chaos_2005,corte}, an experimental system of non-Brownian
colloids displaying a transition to an absorbing state has been studied,
and a model, called ``random organization''\cite{corte}, has been
proposed as a description of the experimental phenomena.

Absorbing states satisfy some model-dependent condition, the specifics of which
do not appear to change the behavior qualitatively. 
The control-parameter space is divided into two regions
- one where absorbing states are achieved for any typical initial
state, the other where absorbing states, although possible, are not
attained, resulting in a non-equilibrium steady state with a non-zero
average number of ``active'' particles which violate the stipulated
condition. These regions are separated by a critical line, and the
transition between the two phases displays characteristics of a continuous
phase transition, with a diverging correlation length. This phenomenology is common to all absorbing state
models.

In this Letter we shall study the ``order'' that
develops in several absorbing state models as the system approaches
the critical point from within the absorbing phase. Remarkably, even
though the initial states are random, the dynamics are random, and
there is an enormous number of random absorbing states, those absorbing
states actually attained by the system are special, with greatly diminshed
density fluctuations. As we approach criticality, these fluctuations
scale differently from the usual random system - they are, in fact,
\emph{hyperuniform}\cite{torquato_local_2003}. Hyperuniform systems
have recently attracted interest due to their relation to photonic
band-gap materials\cite{BandGap1,BandGap2}. Thus, the setup in \cite{corte,pine_chaos_2005}
may provide an efficient method to create such materials in bulk. 

For a system in $d$ dimensions, we characterize the density fluctuations
in a region of volume $\ell^{d}$ as a function of $\ell$ by 
\begin{equation}
\sigma^{2}\left(\ell\right)\,\equiv\,\left\langle \rho^{2}\left(\ell\right)\right\rangle -\left\langle \rho\left(\ell\right)\right\rangle ^{2}\,\propto\;\ell^{-\lambda}\label{eq:Varell}
\end{equation}
Random systems, such as a Poisson process, have the standard ``$\:\sqrt{N}$
'' fluctuations; this corresponds to $\lambda=d$. When $\lambda>d$,
the particles are distributed more uniformly; such distributions are
termed hyperuniform\cite{torquato_local_2003}. For example, a crystal
whose atoms are displaced by small random amounts from their lattice
sites will have $\lambda=d+1$. We find that for the models we study,
$\lambda>d$ for critical absorbing states. In all cases, the initial
configuration is random with no correlations ($\sigma^{2}\left(\ell\right)\propto\ell^{-d}$),
hence any anomalous value of the exponent $\lambda$ is due to the
dynamics. This behavior should be contrasted with equilibrium models,
like the Ising model, whose fluctuations are enhanced at the critical
point, with $\lambda<d$. 

If the system is not exactly critical, we find a crossover from hyperuniform
behavior at short scales to random behavior at large scales; this
enables the definition of a static correlation length $\xi$ which
diverges as the critical line is approached. This behavior is common
to all the absorbing state models we examined, with exponents that
appear to be universal. We note that this correlation length is different
in character and origin from that typically studied for such systems,
which is defined in the active phase.

The systems we shall study are representatives of the conserved directed
percolation (``Manna'') universality
class\cite{hinrichsen_non-equilibrium_2000,Non-Equilibrium_Book}.
In all cases, the particles in a given configuration are distinguished
into ``active'' and ``inactive'' particles. At each time step,
the active particles, if there are any, are displaced randomly with
a bounded distribution. Particle displacements are simultaneous,
and the dynamics continues until there are no more active particles
(this is the absorbing phase) or until a steady state sets in (with
a finite active density). Although it is not essential, we employ
periodic boundary conditions and all the data we present is averaged
over 50-100 realizations.

The various models differ in the criteria defining which of the particles
are active and what sort of random displacements will be performed.
Whatever their details, the phase transitions are characterized by
a set of universal critical exponents which depend only on the dimensionality.
The models we consider in this Letter are:\\
1) \textbf{Conserved Lattice Gas (CLG)} (2D and 3D)\cite{CLG}: Particles
are placed randomly on a $d$-dimensional cubic lattice of volume
$L^{d}$ with density $\rho$, such that each lattice site is occupied
by at most one particle. Particles with adjacent neighbors are considered
active. At each time step, each active particle moves to a random
empty neighboring site. If several active particles attempt to move
to the same empty site, only one of these moves is allowed; the other
particles remain where they were. %
\footnote{We note that the 1D version of this model does not belong to the same
universality class; the critical density is $\rho_{c}=0.5$, hence
the configuration must be periodic. %
}\\
2) \textbf{Manna Model} (1D)\cite{hinrichsen_non-equilibrium_2000}:
Particles are distributed randomly on a one-dimensional lattice of
length $L$ with density $\rho$, allowing multiple occupancy. If
there are more than two particles on a site, they are considered active,
and each of them independently moves randomly to the neighboring site
to the left or the right. \textbf{}\\
\textbf{3) Random Organization }(1D)\textcolor{black}{: $N$ particles
of unit diameter are distributed at random positions in a one dimensional
segment of length $L$. Overlapping particles are considered active,
and at every time step, each is given a random displacement, distributed
uniformly in $\left[-\epsilon,\epsilon\right]$ . In this work, we
have taken $\epsilon=0.25$. }\\
4) \textbf{Random Organization }(2D)\cite{corte}: Particles with
diameter $d$ are distributed uniformly in an $L\times L$ square,
with a covering fraction\textcolor{black}{{} $\phi=\frac{N}{L^{2}}\frac{\pi d^{2}}{4}$.
}The system is then sheared in the $x$ direction with strain amplitude
$\gamma$. If in this process two particles collide, they are deemed
active%
\footnote{Note that this is exactly equivalent to surrounding each particle
by an envelope of a particular shape and asking whether the envelopes
overlap.%
} and are given a random displacement whose magnitude is drawn uniformly
from the range $\left[0,d/2\right]$, with a uniformly chosen angle.
$\gamma$ and $\phi$ are the control parameters of the model, and he critical value of $\gamma$ depends on $\phi$. 

We are particularly interested in the spatial distribution of particles,
and we measure the variance of the density $\sigma^{2}\left(\ell\right)$
in a hypercube of volume $\ell^{d}$ to see how it scales with $\ell$,
as in Equation \ref{eq:Varell}. The asymptotic scaling is related
to  large scale correlations and is not sensitive to short range
fluctuations. 
Whenever $\lambda\neq d$, long range correlations are present in
the system%
\footnote{We note two limiting cases: (a) Poisson distributed particles: $\sigma^{2}\left(\ell\right)\sim\ell^{-d}$,
and (b) particles in a perfect lattice: $\sigma^{2}\left(\ell\right)\sim\ell^{-(d+1)}$.%
}. To see this, consider the variance of the total number of particles
in an infinite system:

\begin{equation}
\left\langle N^{2}\right\rangle -\left\langle N\right\rangle ^{2}=\left\langle N\right\rangle +\int d^{d}r_{1}\int d^{d}r_{2}h\left(r_{1},r_{2}\right)
\end{equation}
where $h\left(r_{1},r_{2}\right)=\left\langle \rho\left(r_{1}\right)\rho\left(r_{2}\right)\right\rangle -\left\langle \rho\left(r_{1}\right)\right\rangle \left\langle \rho\left(r_{2}\right)\right\rangle $
is the two-point correlation function. If $h$ is translationally
invariant, $h\left(r_{1},r_{2}\right)=h\left(r_{1}-r_{2}\right)$,
and since the system volume is infinite, we get 
\begin{equation}
\frac{\left\langle N^{2}\right\rangle -\left\langle N\right\rangle ^{2}}{\left\langle N\right\rangle }=1+\frac{1}{\rho}\int d^{d}rh\left(r\right).\label{eq:VarN_corr}
\end{equation}
If $h\left(r\right)$ decays exponentially with a finite correlation
length, this gives $\lambda=d$, while long-range correlations (power
law decay of $h$) allow for different values of $\lambda$. For a
hyperuniform system, the left hand side of Equation \ref{eq:VarN_corr}
must vanish in the infinite system limit, so we must have $\int d^{d}rh\left(r\right)=-\rho$. 

If now, one considers a finite (large) hyperuniform system, then

\begin{align}
\frac{\left\langle N^{2}\right\rangle -\left\langle N\right\rangle ^{2}}{\left\langle N\right\rangle } & \simeq1+\frac{1}{\rho}A\int_{0}^{\ell}drh\left(r\right)r^{d-1}\\
 & =-\frac{1}{\rho}A\int_{\ell}^{\infty}drh\left(r\right)r^{d-1},\nonumber 
\end{align}
where $A$ is the angular integral. Differentiating with respect to
$\ell$ and comparing to Equation \ref{eq:Varell}, we get that
$h\left(r\right)\sim-\, r^{-\lambda}\label{eq:corr_fn}$.
Hence long-range \textcolor{black}{negative} correlations characterize
hyperuniform systems. The converse, however, is not necessarily true;
long-range correlations are not sufficient for a system to be hyperuniform:
For hyperuniform systems, $\int d^{d}rh\left(r\right)$ is finite,
which is different from many other critical systems, such as the Ising
model, where the integral over the two-point (magnetization) correlation
function diverges.

Power law correlations in continuous phase transitions often imply a diverging length scale. 
Here, this is manifested through motion of particles on a diverging length scale.
To see this, consider a (large) region of volume $\ell^{d}$ in a
much larger system. The number of particles in this volume in the
initial configuration is of order $\rho\ell^{d}\pm B\ell^{d/2}$ where
$B$ is a function of the density. On the other hand, in the final
configuration the number of particles is of the order $\rho\ell^{d}\pm C\ell^{d-\lambda/2}$,
where $\lambda>d$. Thus, for large $\ell$, of order $\ell^{d/2}$
particles will need to move. This must be true at all scales, 
implying that hyperuniformity can be obtained only by moving particles
over large length scales. 

We now discuss the numerical results for the different models, starting
with the 2D CLG. 
\begin{figure}[h]
\includegraphics[scale=0.6]{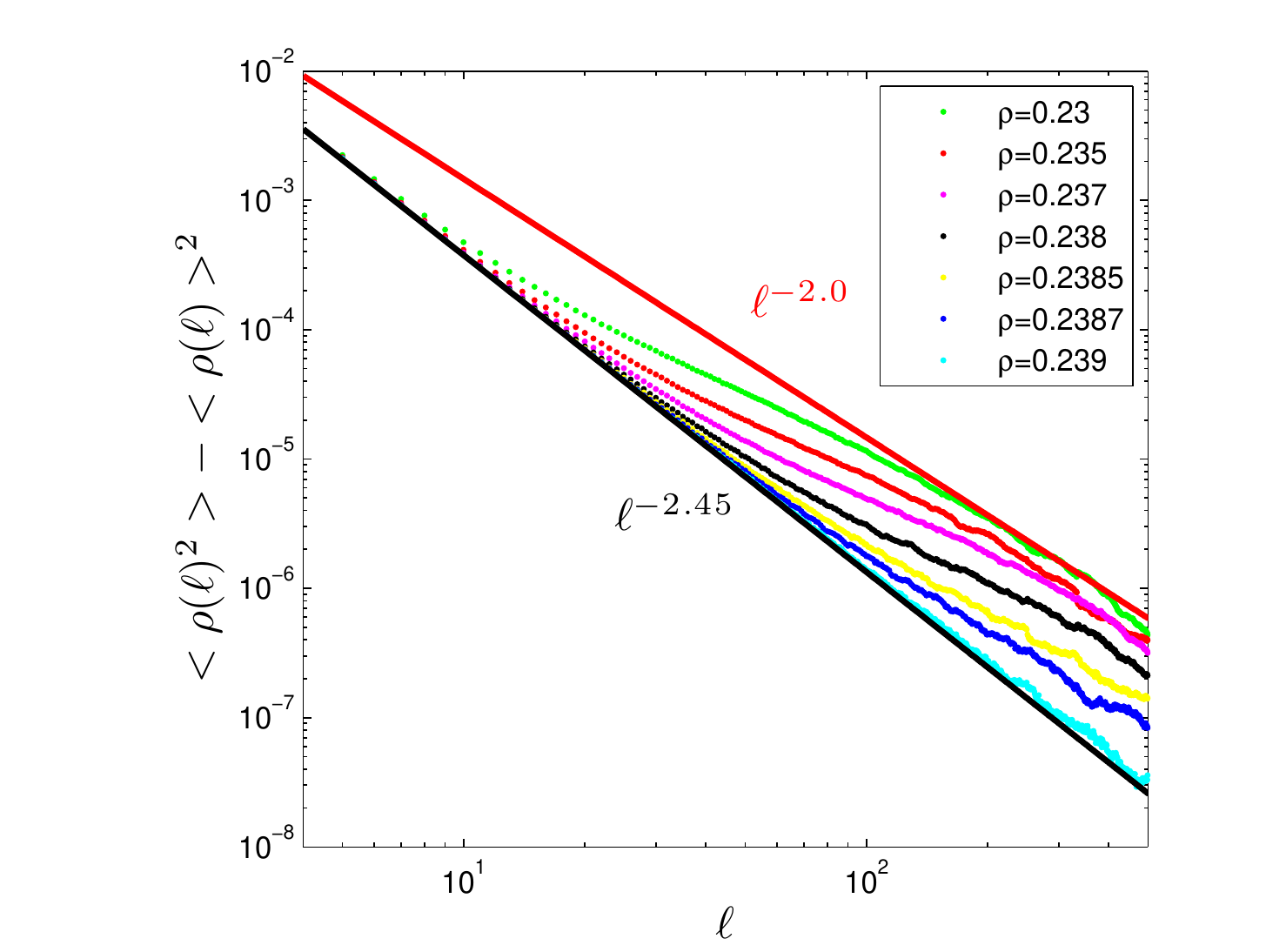} \caption{Density fluctuations in an $\ell\times\ell$ box for the
2D CLG for some different densities.  Here
$L=1000$ and $\rho_{c}\simeq0.2391$. \label{fig:CLG}}
\end{figure}
In Figure \ref{fig:CLG} we plot $\sigma^{2}\left(\ell\right)$ \emph{vs.}
$\ell$ for several densities below the critical density $\rho_{c}\simeq0.2391$.
For $\rho<\rho_{c}$ we find at short distances that $\sigma^{2}\left(\ell\right)$
scales as $\ell^{-\lambda}$ with $\lambda=2.45\pm0.03$, going over,
at larger distances, to a $\ell^{-2}$ decay. \textcolor{black}{This
enables us to define a crossover length $\xi_{\times}$ on the absorbing
side of the transition.} As $\rho$ approaches $\rho_{c}$, $\xi_{\times}$
grows, until finally at $\rho_{c}$, it diverges, and the second scaling region disappears. 

\begin{figure}[h]
\includegraphics[scale=0.6]{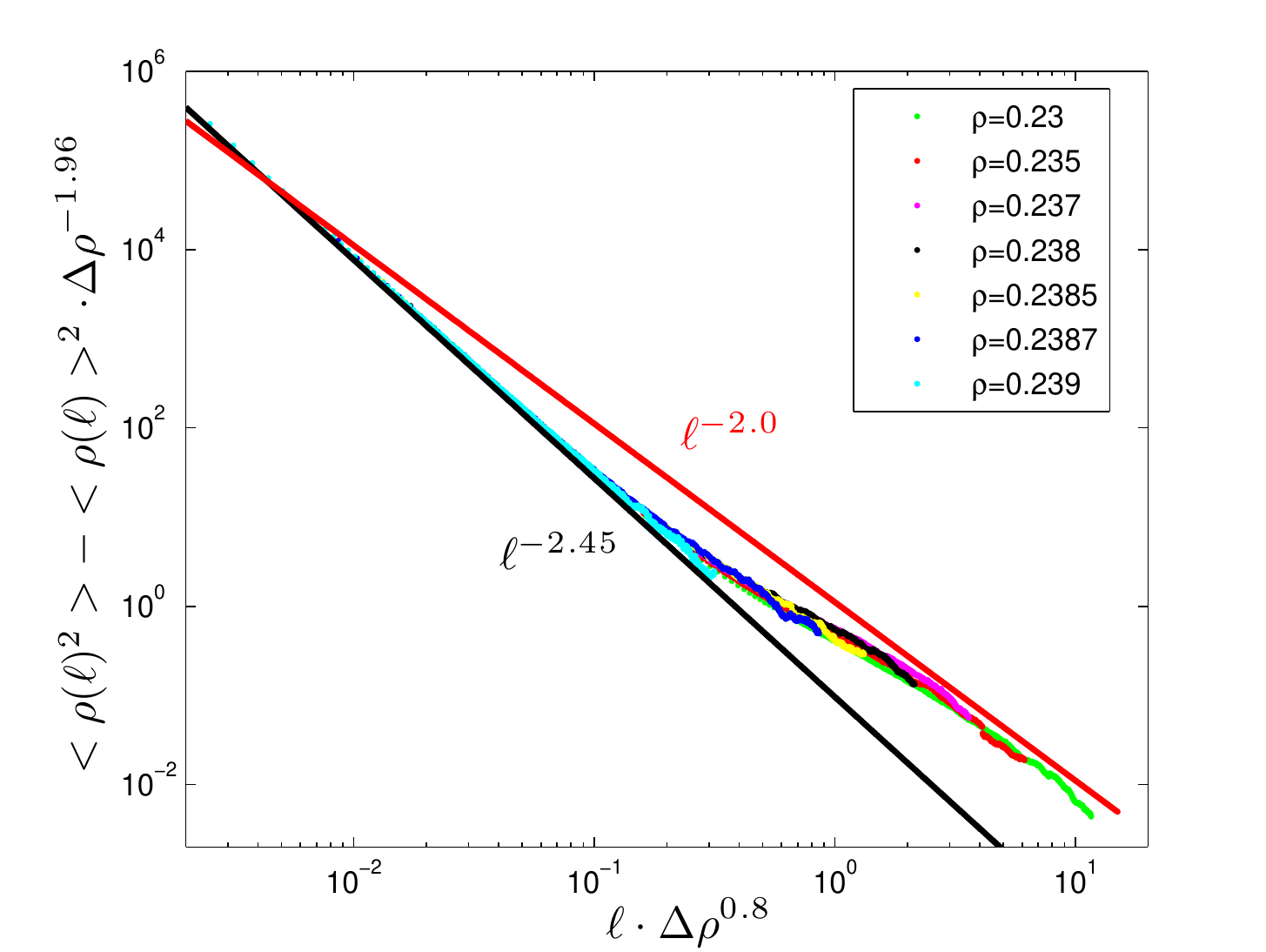} \caption{Data collapse for the mean square density fluctuations for the 2D
CLG model. This supports the contention
that $\xi_{\times}$ has the same scaling with density as does the
active-active correlation function. Here $L=1000$, $\rho_{c}\simeq0.2391$.
\label{fig:CLG_collapse}}
\end{figure}

To check if $\xi_{\times}$ has the same dependence as the active
site-active site correlation length\cite{hinrichsen_non-equilibrium_2000}
which diverges as $\xi_{AA}\propto\left|\rho-\rho_{c}\right|^{-\nu_{\perp}}\equiv\left|\Delta\rho\right|^{-\nu_{\perp}}$,
we rescale $\ell$ by $\left|\Delta\rho\right|^{-\nu_{\perp}}$, where
$\nu_{\perp}\simeq0.8$ in 2D\cite{Lubek}. A true data collapse is
not quite possible since there are two scaling regimes. We choose
to match the first scaling regime, so we rescale the $y$ axis by
$\Delta\rho^{-\nu_{\perp}\lambda}$. As seen in Figure \ref{fig:CLG_collapse},
the data collapses rather well, suggesting that $\xi_{\times}$ scales
in the same way as the active-active correlation length, despite the
fact that the former is defined in the absorbing phase, and the latter
in the active phase.

To relate to the experiment of Reference \cite{pine_chaos_2005},
and to test the universality of the exponent $\lambda$, we simulate
the 2D random organization model introduced in \cite{corte}. This
model is intrinsically anisotropic, due to the directionality of the
shear. As in the experiment, we set the covering fraction $\phi$
and vary the strain $\gamma$. 
\begin{figure}
\begin{centering}
\includegraphics[scale=0.6]{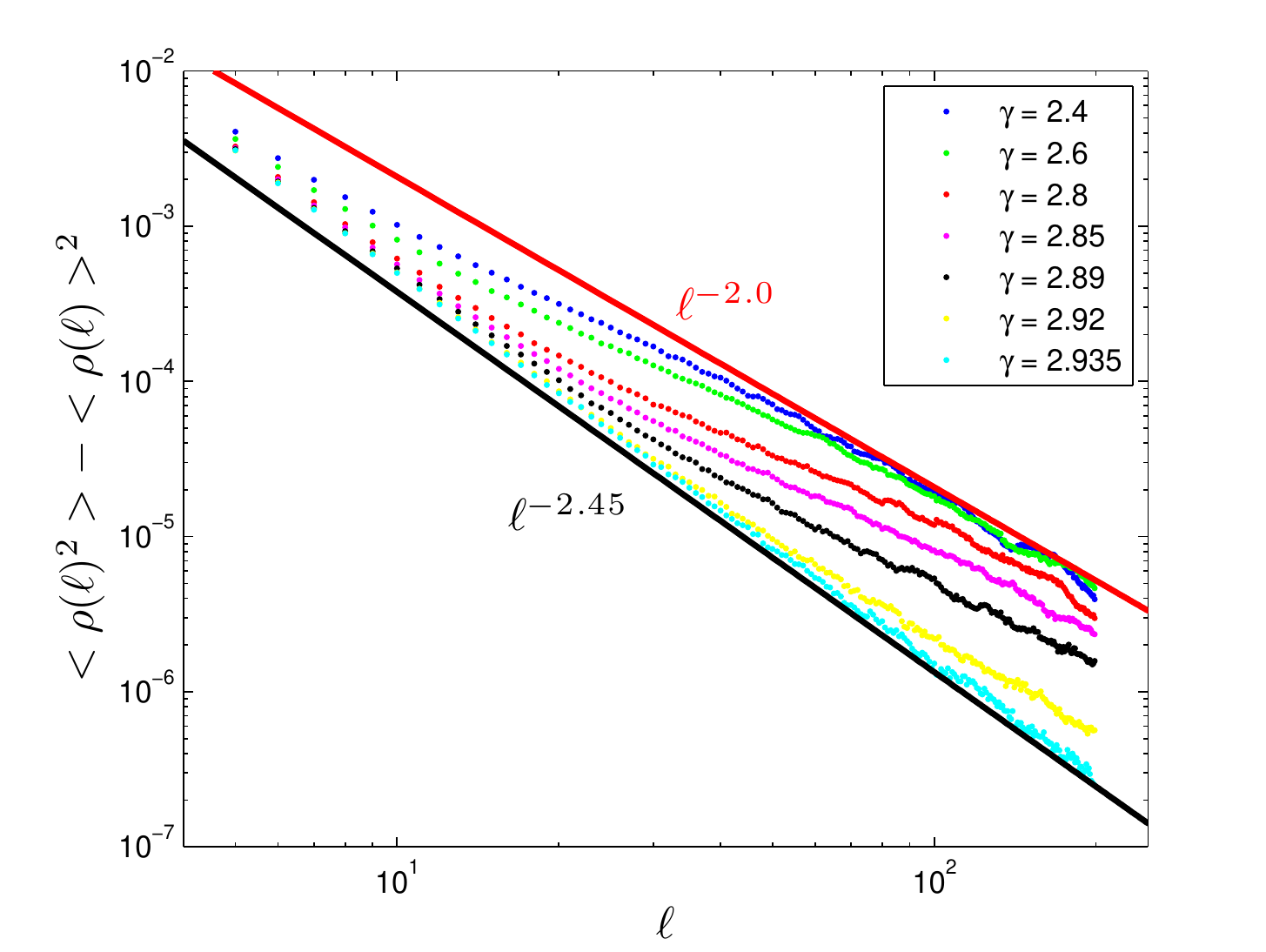}
\par\end{centering}

\caption{Mean square density fluctuations for the 2D random organization model
at various values of strain $\gamma<\gamma_{c}\simeq2.935$, with
$\phi=0.2$ and $L=400$.\label{fig:Shear2d} }
\end{figure}
In Figure \ref{fig:Shear2d}, $\sigma^{2}\left(\ell\right)$ is shown
for $\phi=0.2$ and various values of the strain. Phenomenology similar
to that of the CLG is seen, with the same value of the exponent $\lambda$,
suggesting that indeed $\lambda$ is universal despite the anisotropy
of the system. This is in agreement with reference \cite{Ramaswamy}
which asserts that anisotropy should not change the universality class
of the system.

In addition to the two-dimensional systems above, we have also simulated
the 1D Manna model and the 1D random organization model of \cite{corte},
as well as the 3D CLG. We find that in both one-dimensional models,
$\lambda_{1d}\simeq1.425\pm0.025$, as shown 
\begin{figure}
\begin{centering}
\includegraphics[scale=0.6]{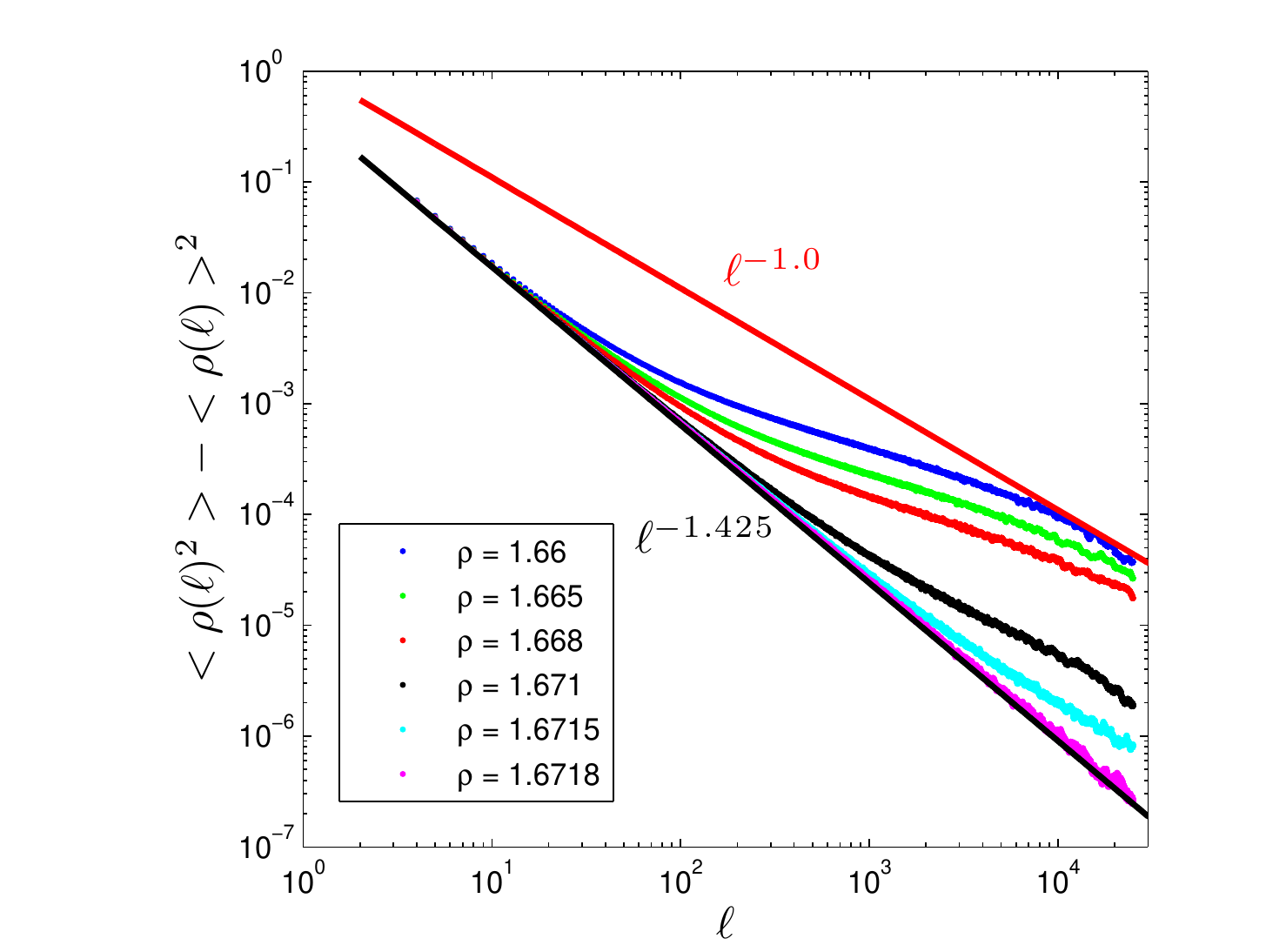}
\par\end{centering}

\caption{Density fluctuations in the 1D Manna model as a function of length
$\ell$. Here $L=100000$ and $\rho<\rho_{c}\simeq1.6718$.
\label{fig:One-dimensional-Manna}}
\end{figure}
\begin{figure}
\begin{centering}
\includegraphics[scale=0.6]{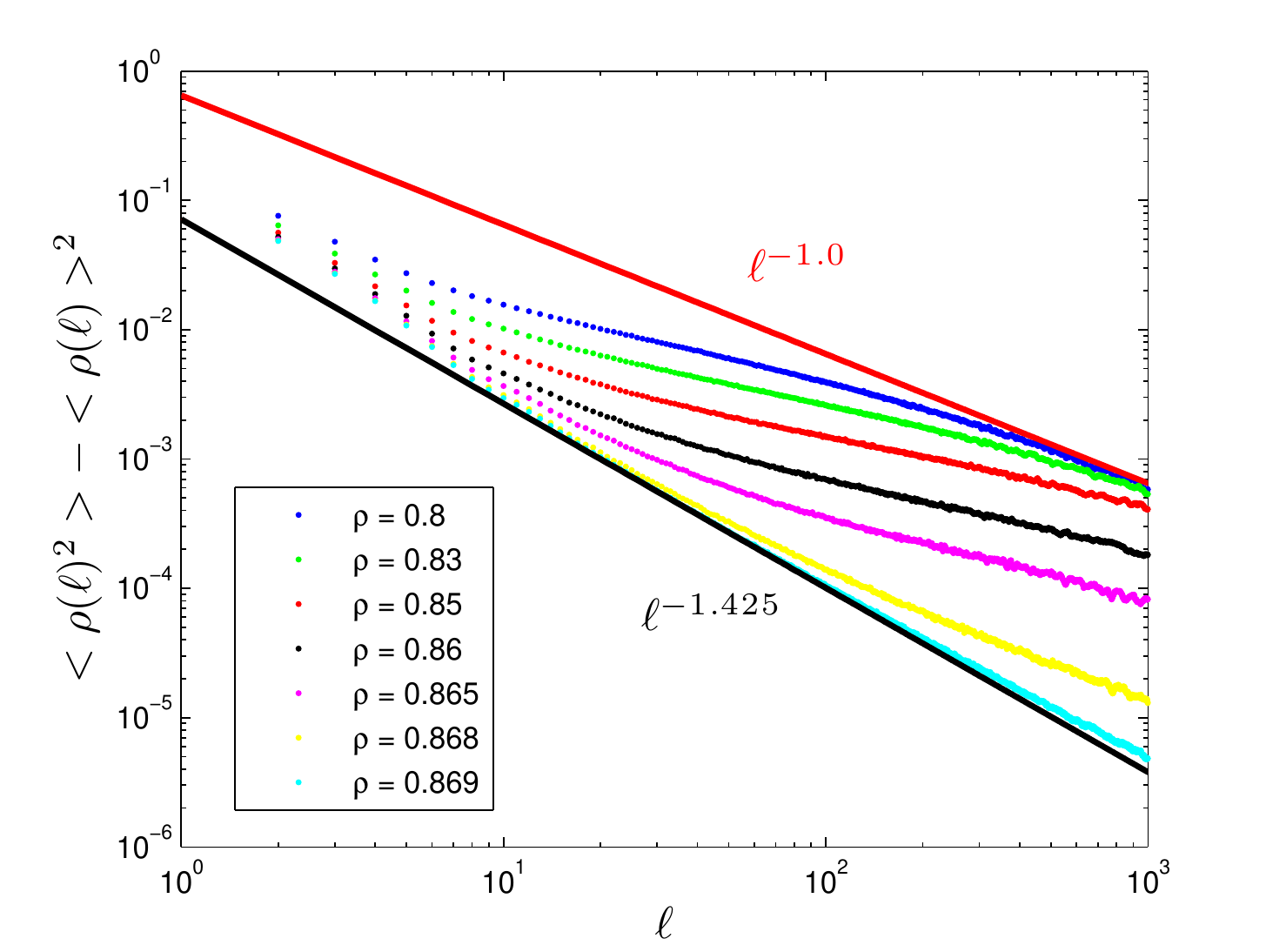}
\par\end{centering}

\caption{Density fluctuations as a function of $\ell$ for 1D random
organization. Here, $L=10000$ and $\rho<\rho_{c}\simeq0.8692$. \label{fig:Shear_1d}}
\end{figure}
in Figures \ref{fig:One-dimensional-Manna} and \ref{fig:Shear_1d},
while for the 3D CLG, $\lambda_{3d}\simeq3.24\pm0.07$, as discussed
below.

\textcolor{black}{The picture that emerges from these results is that
wherever the dynamics act, they smooth out density fluctuations in
such as way as to become hyperuniform}%
\footnote{\textcolor{black}{This crossover is clearly seen for the CLG and Manna
models. For random organization, the finite size of the particle (or,
equivalently, the applied strain) obscures this at very small scales.
It becomes more evident, however, close to the critical point, as
seen in Figures \ref{fig:Shear2d} and \ref{fig:Shear_1d}.}%
}\textcolor{black}{. When the system is in the absorbing phase (say,
at low densities), the dynamics take place over regions of size $\xi_{\times}$,
so the system becomes hyperuniform on length scales of this order.
As criticality is approached, the dynamics occur over larger and larger
scales, and $\xi_{\times}$ diverges. For this reason. $\xi_{\times}$
can be regarded as a basic correlation length describing the ordering
of the system. }

Experimentally, hyperuniformity is most readily identified by measuring
the structure factor $S(k)$
\footnote {The structure factor is defined by
$S\left(k\right)=\frac{1}{N}\:\left\langle \:\left|\sum_{j=1}^{N}e^{ikr_{j}}\right|^{2}\:\right\rangle$ },
which, for small values of $k$, scales as $S(k)=Ak^{\lambda-d}+N\delta(k)$:

\begin{align}
S\left(k\right) & =1+\frac{1}{\rho}\int d^{d}r\left\langle \rho\left(r\right)\rho\left(0\right)\right\rangle e^{-ikr}\\
 & =1+\frac{1}{\rho}\int d^{d}rh\left(r\right)e^{-ikr}+N\delta\left(k\right)\nonumber 
\end{align}
With $h\left(r\right)\propto r^{-\lambda}$ , the result follows.  Disregarding the delta function, $S\left(k\right)\rightarrow0$
as $k\rightarrow0$; this is the signature of hyperuniformity.

\begin{figure}[h]
\includegraphics[scale=0.6]{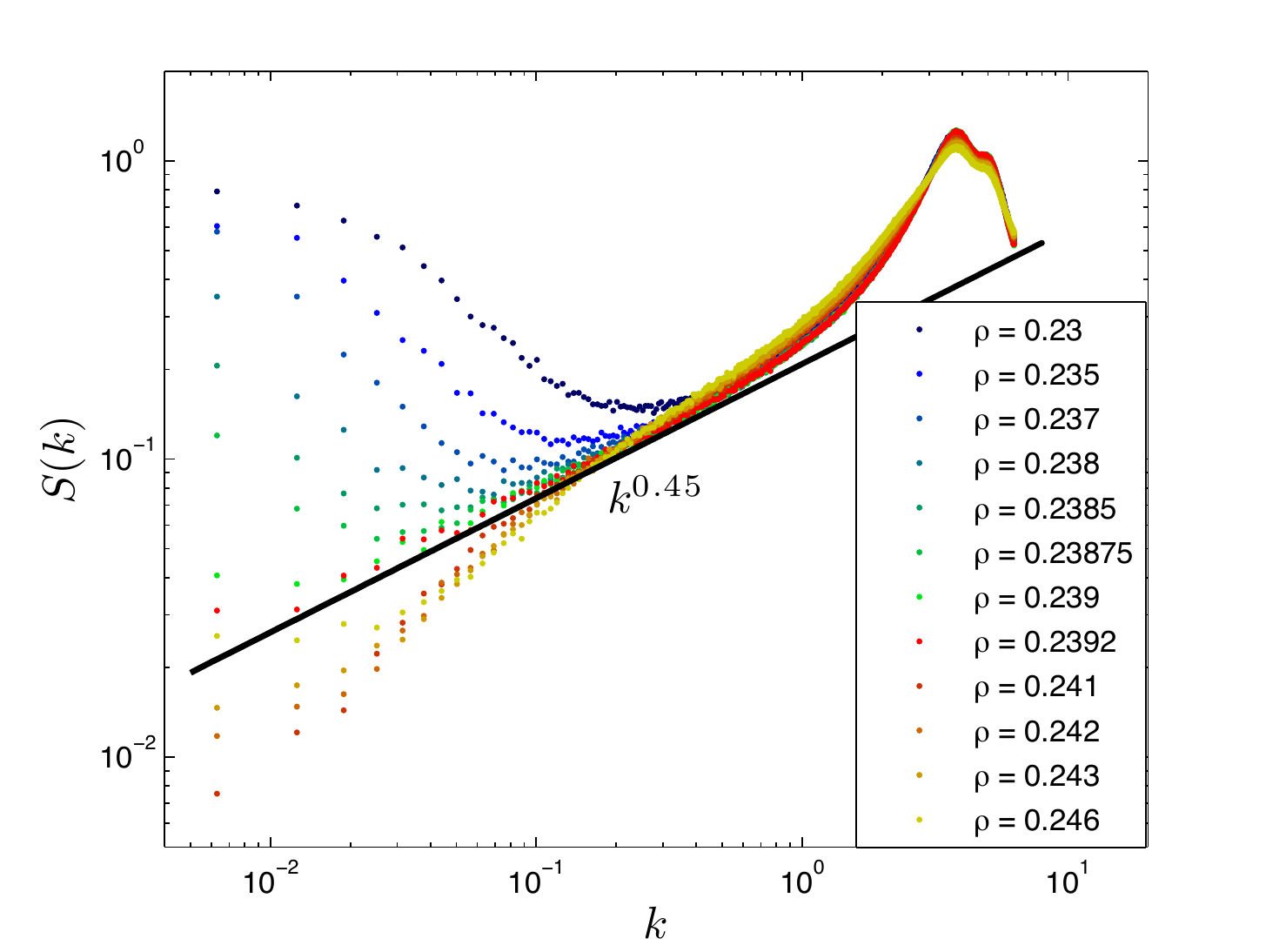} \caption{The structure factor $S(k)$ at small $k$ for the 2D CLG. Here $L=1000$
and $\rho_{c}\simeq 0.2391$. \label{fig:Sq}}
\end{figure}

Let us consider the small $k$ behavior of $S(k)$ for
the 2D CLG (other models are similar). To do this, we first evaluate
$S(k)$ numerically at values of $\vec{k}=\left(n,m\right)\frac{2\pi}{L}$,
with $n,m\mathbb{\,}$ integers%
\footnote{Note that this is equivalent to putting the system under periodic
boundary conditions, which removes the finite size effects near $k=0$.%
}, and then perform an angular average for a given $\left|k\right|=\frac{2\pi n}{L}$
using all the wave vectors in the range $\frac{2\pi n}{L}\leq\left|k\right|<\frac{2\pi\left(n+1\right)}{L}$.
As seen in Figure \ref{fig:Sq}, at the critical point, $S(k)\sim\left|k\right|^{\lambda-d}$
for small $k$, with $\lambda\simeq2.45.$ When $\rho<\rho_{c}$,
$S(k)$ is proportional to $\left|k\right|^{\lambda-d}$
for intermediate values of $k$, with $S(k)\rightarrow\; const$
as $k\rightarrow0$. 

Surprising behavior is observed for $\rho>\rho_{c}$ in the active
region. Here, we measure $S(k)$ from snapshots of the
system, since it includes active particles and is dynamic. Slightly
above the critical density, $S(k)$ appears to drop faster than $\left|k\right|^{\lambda-d}$
for small values of $k$, as shown in Figure \ref{fig:Sq}: fluctuations
are reduced on large scales. We believe that this behavior must get
cut off at large enough length scales when off criticality, with $lim_{k\rightarrow0}\, S\left(k\right)>0$
indicating a finite (perhaps large) correlation length.


%

Last, we observe a relation between the critical exponent $\lambda$
to a known exponent of the Manna universality class. First, we note
that at criticality, the active-active correlation function (measured
in the active phase, of course) goes as 
$c(r)\sim r^{2-d-\eta_{\bot}}$,
where $\eta_{\bot}$ is an exponent whose values are given in Table
\ref{tab:scaling_relation}. As noted above, the
density-density correlation function at criticality scales as $h(r)\sim r^{-\lambda}$.
As shown in Table \ref{tab:scaling_relation}, our data is well fit
in all dimensions by the form 
\begin{equation}
\lambda=d+2-\eta_{\bot}
\end{equation}
which is different from the active-active correlation exponent.

\begin{table}
\begin{centering}
\begin{tabular}{|c|c|c|c|}
\hline 
$d$ & $\lambda$ & $\eta_{\perp}$ & $d+2-\eta_{\perp}$\tabularnewline
\hline 
\hline 
$1$ & $1.425\pm0.025$ & $1.592\pm0.040$ & $1.4076\pm0.040$\tabularnewline
\hline 
$2$ & $2.45\pm0.03$ & $1.541\pm0.025$ & $2.4588\pm0.025$\tabularnewline
\hline 
$3$ & $3.24\pm0.07$ & $1.744\pm0.029$ & $3.2557\pm0.029$\tabularnewline
\hline 
\end{tabular}
\par\end{centering}

\caption{The measured values of $\lambda$ and the proposed scaling relation.
Values for $\eta_{\perp}$ are from reference \cite{Non-Equilibrium_Book}.
The exponent $\lambda$ and its error were measured in 1D for the
Manna model while in 2D and 3D for the CLG model. \label{tab:scaling_relation}}
\end{table}

We conclude by comparing this behavior to that of an equilibrium system.
In equilibrium, particle number fluctuations are related to the isothermal
compressibility $\kappa_{T}\equiv-\frac{1}{V}\left(\frac{\partial V}{\partial P}\right)_{T}$
by
$\left\langle N^{2}\right\rangle -\left\langle N\right\rangle ^{2}=\left\langle N\right\rangle \rho k_{B}T\kappa_{T}$.
If $\kappa_{T}$ is finite and non-zero, the particle number variance
is extensive, yielding $\lambda=d$.
This result is expected 
for finite ranged interactions since there is a finite
energy cost to move a particle to some random location, so that the
density of displaced particles is proportional to the system volume.
Hyperunifomity, therefore, would not be manifested in equilibrium
states of systems with finite-range interactions at finite temperature.

We would like to thank Bill Bialek, Andrea Cavagna, Paul Chaikin,
and David Pine for interesting and informative discussions. DL thanks
the US-Israel Binational Science Foundation (grant 2008483) and the
Israel Science Foundation (grant 1254/12) for support.

\bibliographystyle{aipnum4-1}
\bibliography{Hyper}

\begin{thebibliography}{10}%
\makeatletter
\providecommand \@ifxundefined [1]{%
 \@ifx{#1\undefined}
}%
\providecommand \@ifnum [1]{%
 \ifnum #1\expandafter \@firstoftwo
 \else \expandafter \@secondoftwo
 \fi
}%
\providecommand \@ifx [1]{%
 \ifx #1\expandafter \@firstoftwo
 \else \expandafter \@secondoftwo
 \fi
}%
\providecommand \natexlab [1]{#1}%
\providecommand \enquote  [1]{``#1''}%
\providecommand \bibnamefont  [1]{#1}%
\providecommand \bibfnamefont [1]{#1}%
\providecommand \citenamefont [1]{#1}%
\providecommand \href@noop [0]{\@secondoftwo}%
\providecommand \href [0]{\begingroup \@sanitize@url \@href}%
\providecommand \@href[1]{\@@startlink{#1}\@@href}%
\providecommand \@@href[1]{\endgroup#1\@@endlink}%
\providecommand \@sanitize@url [0]{\catcode `\\12\catcode `\$12\catcode
  `\&12\catcode `\#12\catcode `\^12\catcode `\_12\catcode `\%12\relax}%
\providecommand \@@startlink[1]{}%
\providecommand \@@endlink[0]{}%
\providecommand \url  [0]{\begingroup\@sanitize@url \@url }%
\providecommand \@url [1]{\endgroup\@href {#1}{\urlprefix }}%
\providecommand \urlprefix  [0]{URL }%
\providecommand \Eprint [0]{\href }%
\providecommand \doibase [0]{http://dx.doi.org/}%
\providecommand \selectlanguage [0]{\@gobble}%
\providecommand \bibinfo  [0]{\@secondoftwo}%
\providecommand \bibfield  [0]{\@secondoftwo}%
\providecommand \translation [1]{[#1]}%
\providecommand \BibitemOpen [0]{}%
\providecommand \bibitemStop [0]{}%
\providecommand \bibitemNoStop [0]{.\EOS\space}%
\providecommand \EOS [0]{\spacefactor3000\relax}%
\providecommand \BibitemShut  [1]{\csname bibitem#1\endcsname}%
\let\auto@bib@innerbib\@empty
\bibitem [{\citenamefont {L{\"u}beck}(2004)}]{Lubek}%
  \BibitemOpen
  \bibfield  {author} {\bibinfo {author} {\bibfnamefont {S.}~\bibnamefont
  {L{\"u}beck}},\ }\href@noop {} {\bibfield  {journal} {\bibinfo  {journal}
  {Int. J. Mod. Phys. B}\ }\textbf {\bibinfo {volume} {18}},\ \bibinfo {pages}
  {3977} (\bibinfo {year} {2004})}\BibitemShut {NoStop}%
\bibitem [{\citenamefont {Hinrichsen}(2000)}]{hinrichsen_non-equilibrium_2000}%
  \BibitemOpen
  \bibfield  {author} {\bibinfo {author} {\bibfnamefont {H.}~\bibnamefont
  {Hinrichsen}},\ }\href@noop {} {\bibfield  {journal} {\bibinfo  {journal}
  {Advances in Physics}\ }\textbf {\bibinfo {volume} {49}},\ \bibinfo {pages}
  {815} (\bibinfo {year} {2000})}\BibitemShut {NoStop}%
\bibitem [{\citenamefont {Henkel}, \citenamefont {Hinrichsen},\ and\
  \citenamefont {L{\"u}beck}(2008)}]{Non-Equilibrium_Book}%
  \BibitemOpen
  \bibfield  {author} {\bibinfo {author} {\bibfnamefont {M.}~\bibnamefont
  {Henkel}}, \bibinfo {author} {\bibfnamefont {H.}~\bibnamefont {Hinrichsen}},
  \ and\ \bibinfo {author} {\bibfnamefont {S.}~\bibnamefont {L{\"u}beck}},\
  }\href@noop {} {\emph {\bibinfo {title} {Non-Equilibrium Phase Transitions -
  Volume 1: Absorbing Phase Transitions}}}\ (\bibinfo  {publisher} {Springer},\
  \bibinfo {year} {2008})\BibitemShut {NoStop}%
\bibitem [{\citenamefont {Pine}\ \emph {et~al.}(2005)\citenamefont {Pine},
  \citenamefont {Gollub}, \citenamefont {Brady},\ and\ \citenamefont
  {Leshansky}}]{pine_chaos_2005}%
  \BibitemOpen
  \bibfield  {author} {\bibinfo {author} {\bibfnamefont {D.~J.}\ \bibnamefont
  {Pine}}, \bibinfo {author} {\bibfnamefont {J.~P.}\ \bibnamefont {Gollub}},
  \bibinfo {author} {\bibfnamefont {J.~F.}\ \bibnamefont {Brady}}, \ and\
  \bibinfo {author} {\bibfnamefont {A.~M.}\ \bibnamefont {Leshansky}},\
  }\href@noop {} {\bibfield  {journal} {\bibinfo  {journal} {Nature}\ }\textbf
  {\bibinfo {volume} {438}},\ \bibinfo {pages} {997} (\bibinfo {year}
  {2005})}\BibitemShut {NoStop}%
\bibitem [{\citenamefont {Cort{\'e}}\ \emph {et~al.}(2008)\citenamefont
  {Cort{\'e}}, \citenamefont {Chaikin}, \citenamefont {Gollub},\ and\
  \citenamefont {Pine}}]{corte}%
  \BibitemOpen
  \bibfield  {author} {\bibinfo {author} {\bibfnamefont {L.}~\bibnamefont
  {Cort{\'e}}}, \bibinfo {author} {\bibfnamefont {P.~M.}\ \bibnamefont
  {Chaikin}}, \bibinfo {author} {\bibfnamefont {J.~P.}\ \bibnamefont {Gollub}},
  \ and\ \bibinfo {author} {\bibfnamefont {D.~J.}\ \bibnamefont {Pine}},\
  }\href@noop {} {\bibfield  {journal} {\bibinfo  {journal} {Nature Physics}\
  ,\ \bibinfo {pages} {420}} (\bibinfo {year} {2008})}\BibitemShut {NoStop}%
\bibitem [{\citenamefont {Torquato}\ and\ \citenamefont
  {Stillinger}(2003)}]{torquato_local_2003}%
  \BibitemOpen
  \bibfield  {author} {\bibinfo {author} {\bibfnamefont {S.}~\bibnamefont
  {Torquato}}\ and\ \bibinfo {author} {\bibfnamefont {F.~H.}\ \bibnamefont
  {Stillinger}},\ }\href@noop {} {\bibfield  {journal} {\bibinfo  {journal}
  {Phys. Rev. E}\ }\textbf {\bibinfo {volume} {68}},\ \bibinfo {pages} {041113}
  (\bibinfo {year} {2003})}\BibitemShut {NoStop}%
\bibitem [{\citenamefont {Man}\ \emph {et~al.}(2013)\citenamefont {Man},
  \citenamefont {Florescu}, \citenamefont {Matsuyama}, \citenamefont {Yadak},
  \citenamefont {Nahal}, \citenamefont {Hashemizad}, \citenamefont
  {Williamson}, \citenamefont {Steinhardt}, \citenamefont {Torquato},\ and\
  \citenamefont {Chaikin}}]{BandGap1}%
  \BibitemOpen
  \bibfield  {author} {\bibinfo {author} {\bibfnamefont {W.}~\bibnamefont
  {Man}}, \bibinfo {author} {\bibfnamefont {M.}~\bibnamefont {Florescu}},
  \bibinfo {author} {\bibfnamefont {K.}~\bibnamefont {Matsuyama}}, \bibinfo
  {author} {\bibfnamefont {P.}~\bibnamefont {Yadak}}, \bibinfo {author}
  {\bibfnamefont {G.}~\bibnamefont {Nahal}}, \bibinfo {author} {\bibfnamefont
  {S.}~\bibnamefont {Hashemizad}}, \bibinfo {author} {\bibfnamefont
  {E.}~\bibnamefont {Williamson}}, \bibinfo {author} {\bibfnamefont
  {P.}~\bibnamefont {Steinhardt}}, \bibinfo {author} {\bibfnamefont
  {S.}~\bibnamefont {Torquato}}, \ and\ \bibinfo {author} {\bibfnamefont
  {P.}~\bibnamefont {Chaikin}},\ }\href@noop {} {\bibfield  {journal} {\bibinfo
   {journal} {Opt. Express}\ }\textbf {\bibinfo {volume} {21}},\ \bibinfo
  {pages} {19972} (\bibinfo {year} {2013})}\BibitemShut {NoStop}%
\bibitem [{\citenamefont {Florescu}, \citenamefont {Torquato},\ and\
  \citenamefont {Steinhardt}(2009)}]{BandGap2}%
  \BibitemOpen
  \bibfield  {author} {\bibinfo {author} {\bibfnamefont {M.}~\bibnamefont
  {Florescu}}, \bibinfo {author} {\bibfnamefont {S.}~\bibnamefont {Torquato}},
  \ and\ \bibinfo {author} {\bibfnamefont {P.~J.}\ \bibnamefont {Steinhardt}},\
  }\href@noop {} {\bibfield  {journal} {\bibinfo  {journal} {PNAS}\ }\textbf
  {\bibinfo {volume} {106}},\ \bibinfo {pages} {20658} (\bibinfo {year}
  {2009})}\BibitemShut {NoStop}%
\bibitem [{\citenamefont {Rossi}, \citenamefont {Pastor-Satorras},\ and\
  \citenamefont {Vespignani}(2000)}]{CLG}%
  \BibitemOpen
  \bibfield  {author} {\bibinfo {author} {\bibfnamefont {M.}~\bibnamefont
  {Rossi}}, \bibinfo {author} {\bibfnamefont {R.}~\bibnamefont
  {Pastor-Satorras}}, \ and\ \bibinfo {author} {\bibfnamefont {A.}~\bibnamefont
  {Vespignani}},\ }\href@noop {} {\bibfield  {journal} {\bibinfo  {journal}
  {Phys. Rev. Lett.}\ }\textbf {\bibinfo {volume} {85}},\ \bibinfo {pages}
  {1803} (\bibinfo {year} {2000})}\BibitemShut {NoStop}%
\bibitem [{\citenamefont {Menon}\ and\ \citenamefont
  {Ramaswamy}(2009)}]{Ramaswamy}%
  \BibitemOpen
  \bibfield  {author} {\bibinfo {author} {\bibfnamefont {G.~I.}\ \bibnamefont
  {Menon}}\ and\ \bibinfo {author} {\bibfnamefont {S.}~\bibnamefont
  {Ramaswamy}},\ }\href@noop {} {\bibfield  {journal} {\bibinfo  {journal}
  {Phys. Rev. E}\ }\textbf {\bibinfo {volume} {79}},\ \bibinfo {pages} {061108}
  (\bibinfo {year} {2009})}\BibitemShut {NoStop}%
\end{thebibliography}%
 
\end{document}